\documentclass[onecolumn]{pasj00}
\begin{document}
\SetRunningHead{F. Oktariani, A.\,T. Okazaki and S. Kato}
{Excitation of Trapped Oscillations in  
Disks around Black Holes}
\Received{2009//}
\Accepted{2010/??/??}

\title{Excitation of Trapped g-Mode Oscillations in Warped Disks around Black Holes}

\draft

\author{Finny \textsc{oktariani}} 
\affil{ Department of Cosmoscience, Graduate School of Science, Hokkaido University, Kita-ku, Sapporo 060-0810, Japan}
\email{finny@astro1.sci.hokudai.ac.jp}
\author{Atsuo T. \textsc{okazaki}}
\affil{Faculty of Engineering, Hokkai-Gakuen University, Toyohira-ku, Sapporo 062-8605, Japan }
\and
\author{Shoji \textsc{kato}}
\affil{2-2-2 Shikanodai-nishi, Ikoma-shi, Nara 630-0114, Japan}

\maketitle

\begin{abstract}
In order to study the origin of high-frequency quasi-periodic oscillations observed in X-ray binaries, \citet{Kat04} suggested a 
resonant excitation mechanism of disk oscillations in deformed disks. In this paper, we investigate numerically, following his formulation, whether trapped g-mode oscillations in a warped disk, where the warp amplitude varies with radius, can be 
excited by this mechanism. For simplicity, we adopt Newtonian hydrodynamic equations with relativistic expressions for the characteristic 
frequencies of disks. 
We also assume that the accretion disk is isothermal. 
We find that the fundamental modes of trapped g-mode oscillations with eigenfrequencies close to the maximum of 
epycyclic frequency are excited. The intermediate oscillations found are isolated in a narrow region around the resonance radius. After varying some parameters, we find that the growth rate increases 
as the warp amplitude or the black hole spin parameter increases, while it decreases as the sound speed increases. 
\end{abstract}

\section{Introduction}
\label{sec:intro}
The Rossi X-ray Timing Explorer (RXTE) satellite has detected  high-frequency quasi-periodic oscillations (QPOs) in X-ray binaries.
They are kilohertz (kHz) and hectohertz (hHz) QPOs in neutron-star low-mass X-ray binaries, 
and high-frequency (HF) QPOs ($\geq 100$ Hz) in black-hole X-ray binaries. 
HF QPOs occur at fixed frequencies and the appearance is correlated to the state of the sources, that is, they
appear only in high-luminosity states where $L > 0.1 L_{E}$, 
with $L_{E}$ being the Eddington luminosity. 
They are suggested to be phenomena originating in a strong gravitational field, 
which occur in the innermost region of the relativistic disk. 
HF QPOs are regarded as being a powerful tool to explore the mass and spin of the central black hole, and also to explore the physical states of the innermost region of the accretion disk.

Various models have been proposed to explain HF QPOs. For example, \citet{Abr01} and \citet{Klu01} pointed out the importance of 
resonant processes as being the cause of HF QPOs. 
In their model, HF QPOs are the result of resonant couplings between the vertical and horizontal epicyclic oscillations 
at a particular radius. 
There is, however, uncertainty concerning the excitation process of the oscillation system as a whole. 

\authorcite{Kat04}\ (\yearcite{Kat04, Kat08a, Kat08b}) proposed a model where HF QPOs are regarded to be disk oscillations resonantly excited 
in deformed disks. 
The deformation considered is a warp or an eccentric disk deformation in the equatorial plane. 
An outline of the model is as follows. A non-linear coupling between a disk oscillation 
(hereafter, an original oscillation) and a deformed part of the disk (warp or eccentric deformation) causes some 
forced disk oscillations (hereafter, intermediate oscillations). 
The intermediate oscillations make a resonant coupling with  
the disk, 
and then feedback to the original oscillation. 
Since the nonlinear feedback process involves a resonance, the original oscillation is amplified or damped. 
There are two kinds of resonances, i.e., horizontal resonance (Lindblad resonance) and vertical resonance.
When the deformation is a warp, the resonance that can excite p- and g-mode oscillations is the horizontal resonance (\authorcite{Kat04}\ \yearcite{Kat04, Kat08a, Kat08b}).
Hence, in this paper we consider the horizontal resonance alone.
In the Keplerian disks this resonant process works only when the disk is general relativistic. 
That is, a non-monotonic radial distribution of epicyclic frequency is necessary for the appearance of resonance 
and for trapping of oscillations. 
Recently, \citet{Fer08} generalized \authorcite{Kat04}'s idea on the excitation of trapped g-mode oscillations and made detailed numerical calculations of the 
growth rates of the oscillations in the case of 
rotating black holes. 
Their results indicate that the coupling mechanism can provide an efficient excitation of trapped g-mode oscillations,
provided that the global deformations reach the inner part of the disk with non-negligible amplitude. 

In this paper, we consider nearly the same problem examined by \citet{Fer08},  
following the formulation by \authorcite{Kat08a} \ (\yearcite{Kat08a,Kat08b}).
The purpose of this paper is to make comparison of results obtained from two different approaches, 
\authorcite{Fer08}'s fully numerical approach and \authorcite{Kat08a}'s analytical one. 
We especially consider the case where the disk is warped. 
In section 2, we summarize the basic equations for original oscillations and intermediate oscillations, and 
an analytical expression for growth rate, which we use in later sections. In section 3, we present our numerical results, and section 4 is devoted 
to conclusions and a discussion. A detailed formulation of the nonlinear coupling terms is provided in Appendix.         

\section{Summary of Basic Equations}
\subsection{Original Oscillations}

We assume an unperturbed disk to be steady and axisymmetric with a flow $\textit{\textbf{u}}_{0}$. 
Then, by using a displacement vector $\boldsymbol{\xi}$, 
a weakly nonlinear hydrodynamical equation for adiabatic perturbations is written as (\cite{Ly67})

\begin{equation}
\rho_{0}\frac{\partial^{2}\boldsymbol{\xi}}{\partial t^{2}} + 2\rho_{0} \left( \boldsymbol{u_{0} \cdot \nabla} \right)\frac{\partial \boldsymbol{\xi}}{\partial t}+ \boldsymbol{L(\xi )} = \rho_{0}\boldsymbol{ C(\xi ,\xi )},
\label{eq:Kato1} 
\end{equation} 
where $\textit{\textbf{L}}(\boldsymbol{\xi}$) is a linear Hermitian operator with respect to $\boldsymbol{\xi}$, given by
\begin{eqnarray}
\boldsymbol{L(\xi )} &= &\rho_{0}\left( \boldsymbol{u_{0} \cdot \nabla} \right)\left( \boldsymbol{u_{0} \cdot \nabla} \right) \boldsymbol{\xi} + \rho_{0}\left( \boldsymbol{\xi \cdot \nabla} \right)(\boldsymbol{\nabla} \psi_{0})+ \boldsymbol{\nabla} \left[(1-\Gamma_{1}) p_{0}\mathrm{div} \boldsymbol{\xi} \right] \nonumber \\
&& - p_{0} \boldsymbol{\nabla} (\mathrm{div} \boldsymbol{\xi} ) - \boldsymbol{\nabla} \left[ \left( \boldsymbol{\xi \cdot \nabla} \right) p_{0} \right] + \left( \boldsymbol{\xi \cdot \nabla} \right) (\boldsymbol{\nabla} p_{0}),
\label{eq: Kato2} 
\end{eqnarray} 
where $\rho_{0}$(\textit{\textbf{r}}) and $p_{0}$(\textit{\textbf{r}}) are the density and pressure 
in the unperturbed state, respectively, $\Gamma_{1}$ is the barotropic index specifying the linear part of the relation 
between Lagrangian variations $\delta p$ and $\delta \rho $, and $\psi_{0}$ is a general-relativistic gravitational 
potential. 
The right-hand side of equation (\ref{eq:Kato1}) represents the weakly nonlinear terms. 
For $\Gamma_{1} = 1$ (isothermal perturbations), 
it is given by (\cite{Kat04}) 
\begin{equation}
\rho_{0}\boldsymbol{C(\xi ,\xi )} = - \frac{1}{2} \rho_{0} \xi_{i} \xi_{j} \frac{\partial^{2}}{\partial r_{i} \partial r_{j}}(\boldsymbol{\nabla} \psi_{0}) - \frac{\partial}{\partial r_{i}} \left( p_{0} \frac{\partial \xi_{i}}{\partial r_{j}} \boldsymbol{\nabla} \xi_{j} \right).
\label{eq:Kato3}
\end{equation}

We write the displacement vector $\boldsymbol{\xi}$ in the form of a normal mode, 
\begin{equation}
\boldsymbol{\xi}(r,t) = \hat{\boldsymbol{\xi}}(r,z) \exp \left[ i(\omega t - m\varphi ) \right] ,
\label{eq:disp}
\end{equation}
where $\omega$ is the frequency and $m$ is the azimuthal wavenumber. 
In the case where the disk is isothermal in the vertical direction and the perturbations have a short radial wavelength
compared with the radial characteristic length of the unperturbed disks, we can separate $\boldsymbol{\xi}$ 
into $\textit{r}$-, $\varphi$-, and $\textit{z}$-components as (\cite{Oka87})  

\begin{equation}
\hat{\xi}_{r} (r,z) = \breve{\xi}_{r,n} (r) \mathcal{H}_{n} (z/H),  
\label{eq: Kato20} 
\end{equation} 
\begin{equation}
\hat{\xi}_{\varphi} (r,z) = \breve{\xi}_{\varphi ,n} (r) \mathcal{H}_{n} (z/H),
\label{eq: Kato21} 
\end{equation} 
\begin{equation}
\hat{\xi}_{z} (r,z) = \breve{\xi}_{z,n} (r) \mathcal{H}_{n-1} (z/H),  
\label{eq: Kato22} 
\end{equation} 
where $n$ is a non-negative integer characterizing the number of node(s) of the oscillation in the vertical direction and $\mathcal{H}_{n}$ is the Hermite Polynomial of argument $z/H$, with $H$ being the vertical scale-height of the disk, which is given by $H = c_{s}/\Omega_{\perp} $, where $c_{s}$ is the sound speed and $\Omega_{\perp}$ is the vertical epicyclic frequency.

We express the $r$-, $\varphi$-, and $z$- components of the homogeneous parts of equation (\ref{eq:Kato1}) as 
 
\begin{equation}
\left[ -(\omega - m\Omega)^{2} + \kappa^{2} -4\Omega^{2} -c_{s}^{2} \frac{d^{2}}{dr^{2}} \right] \breve{\xi}_{r,n} - i2\Omega (\omega - m\Omega) \breve{\xi}_{\varphi ,n} + \Omega_{\perp}^{2} H \frac{d\breve{\xi}_{z,n}}{dr} = 0 ,
\label{eq: Kato23} 
\end{equation} 
\begin{equation}
-(\omega - m\Omega)^{2} \breve{\xi}_{\varphi ,n}+ i2\Omega (\omega - m\Omega) \breve{\xi}_{r,n} = 0 ,
\label{eq: Kato24} 
\end{equation} 
\begin{equation}
\left[ -(\omega - m\Omega)^{2} + n\Omega_{\perp}^{2} \right] \breve{\xi}_{z,n} -  n\Omega_{\perp}^{2} H \frac{d\breve{\xi}_{r,n}}{dr} = 0 . 
\label{eq: Kato25} 
\end{equation} 
Note that these equations are the same as equations (23) - (25) of \citet{Kat08b}. 
From these equations we have the basic equation describing the original oscillation (\cite{Oka87}): 
\begin{equation}
\frac{d^{2}\breve{\xi}_{r,n}}{dr^{2}} = \frac{\left[ (\omega -m\Omega)^{2} - \kappa^{2} \right] \left[ n\Omega _{\perp}^{2} - (\omega -m\Omega)^{2} \right]}{\Omega _{\perp}^{2}H^{2} (\omega -m\Omega)^{2}} \breve{\xi}_{r,n}, 
\label{eq:origosc} 
\end{equation} 
where $\Omega $ is the (Keplerian) angular frequency of disk rotation, and $\kappa$ is the horizontal 
epicyclic frequency, respectively. In the general relativistic potential around a rotating black hole, 
these frequencies are given by (e.g., \cite{Kat90})
\begin{equation}
\Omega = (r^{3/2}+a)^{-1},  
\label{eq: Fer15} 
\end{equation} 
\begin{equation}
\kappa = \Omega \sqrt{1-\frac{6}{r}+\frac{8a}{r^{3/2}}-\frac{3a^{2}}{r^{2}}},
\label{eq: Fer16} 
\end{equation} 
\begin{equation}
\Omega_{\perp} = \Omega \sqrt{1-\frac{4a}{r^{3/2}}+\frac{3a^{2}}{r^{2}}},
\label{eq: Fer17} 
\end{equation} 
where $a \ (-1\leq a\leq 1)$ is the central star spin parameter, 
and $r$ is in units of the gravitational radius $R_{g}$, where $R_{g} = GM/c^{2}$ with $M$ being the mass of the central star.  

\subsection{Warp and Intermediate Oscillations}

As mentioned in section 1, our resonant excitation model for the high-frequency QPOs assumes the disk to be warped or 
to have an eccentric deformation. 
In this paper, we consider the case where the disk is warped. The Lagrangian displacement associated with the deformation, $\boldsymbol{\xi}^{W} (\boldsymbol{r},t)$, is given by 
\begin{equation}
\boldsymbol{\xi}^{W} (\boldsymbol{r},t) = \exp (-i m_{W}\varphi)\hat{\boldsymbol{\xi}}^{W} (r,z),
\label{eq:Kato5}
\end{equation}
where $m_{W} = 1$.
\begin{figure}
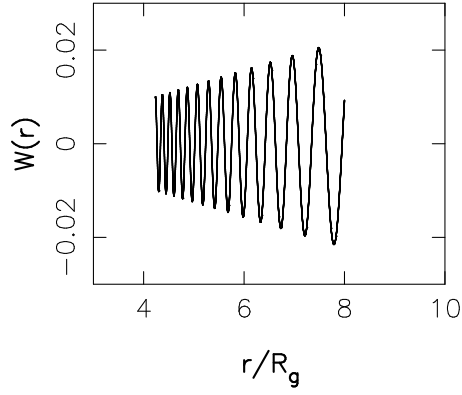

       \centering
         \FigureFile(6cm,6cm){warpgrafba5.ps}
    
		\caption{Warp function $W(r)$ for $c_{s}= 10^{-3}$, $W_{0}= 0.01$, and $a=0.5$. }
    \label{fig:warpsol5}
\end{figure}
The warp solution is described by $u_{r}^{W}=-W(r)\Omega z$, $u_{\varphi}^{W}= iW(r)z d(r\Omega)/dr$, and $u_{z}^{W}= W(r)\Omega r$, where $u_{r}^{W}$, $u_{\varphi}^{W}$, and $u_{z}^{W}$ are the Eulerian perturbation of the velocity field and $W(r)$ is the solution of 

\begin{equation}
\frac{d}{dr} \left(\frac{\Omega^{2}}{\kappa^{2}-\Omega^{2}}\frac{dW}{dr} \right) + \frac{1}{r}\frac{dW}{dr} = \frac{\Omega^{2}-\Omega_{\bot}^{2}}{c_{s}^{2}} W
\label{eq:solwarp} 
\end{equation} 
(\cite{Pap95}; see also \cite{Fer08}). As the boundary condition, we specify the value of warp amplitude at the inner radius,$W_{0}$, and set $dW/dr = 0$ there, following \citet{Fer08}. Here and hereafter, we take the inner radius to be the radius of the Innermost Stable Circular Orbit (ISCO), 
which is the radius where the horizontal epicyclic frequency, $\kappa$, goes to zero. Figure \ref{fig:warpsol5} shows a typical warp function $W(r)$, where $c_{s}= 10^{-3}$, $W_{0}= 0.01$, and $a=0.5$. 

Using the warp function $W(r)$, we can write $\hat{\boldsymbol{\xi}}^{W}$ as 
\begin{equation}
\hat{\xi}_{r}^{W} = - i W(r) z = -i W(r) H \mathcal{H}_{1} (z/H), 
\label{eq: warp1} 
\end{equation} 
\begin{equation}
 \hat{\xi}_{\varphi}^{W} = - W(r) z = -W(r) H \mathcal{H}_{1} (z/H), 
\label{eq: warp2} 
\end{equation} 
\begin{equation}
\hat{\xi}_{z}^{W} = i W(r) r = i W(r) r \mathcal{H}_{0} (z/H). 
\label{eq: warp3} 
\end{equation} 

Nonlinear coupling between the original oscillation, $\boldsymbol{\xi}$, characterized by $(\omega,m$) and the deformation $\boldsymbol{\xi}^{W}$ characterized by $(0,1)$ introduces intermediate oscillations, $\boldsymbol{\xi}^{\mathrm{int}}$, which can be expressed as
\begin{equation}
\boldsymbol{\xi}^{\mathrm{int}}_{\pm} (\boldsymbol{r},t) = \exp \left[ i(\omega t - \tilde{m} \varphi) \right] \hat{\boldsymbol{\xi}}^{\mathrm{int}}_{\pm} (r,z), 
\label{eq: Kato7} 
\end{equation} 
where $\tilde{m}=m+1$ or $m-1$. Here $\hat{\boldsymbol{\xi}}^{\mathrm{int}}_{+}$ represents the intermediate oscillations with $\tilde{m} = m + 1$ resulting from the coupling between $\hat{\boldsymbol{\xi}}$ and $\hat{\boldsymbol{\xi}}^{W}$, while $\hat{\boldsymbol{\xi}}^{\mathrm{int}}_{-}$ represents those with $\tilde{m} = m - 1$ resulting from the coupling between $\hat{\boldsymbol{\xi}}$ and $\hat{\boldsymbol{\xi}}^{W\ast}$, where the asterix represents the complex conjugate. To consider these various coupling cases separately, we write $\hat{\boldsymbol{\xi}}^{\mathrm{int}}_{\pm}$ in the following form:  
\begin{equation}
\hat{\xi}^{\mathrm{int}}_{r,\pm} (r,z) = \breve{\xi}^{\mathrm{int}}_{r,\pm ,\tilde{n}} (r) \mathcal{H}_{\tilde{n}} (z/H),  
\label{eq: Kato26} 
\end{equation} 
\begin{equation}
\hat{\xi}^{\mathrm{int}}_{\varphi ,\pm} (r,z) = \breve{\xi}^{\mathrm{int}}_{\varphi ,\pm ,\tilde{n}} (r) \mathcal{H}_{\tilde{n}} (z/H), 
\label{eq: Kato27} 
\end{equation} 
\begin{equation}
\hat{\xi}^{\mathrm{int}}_{z,\pm} (r,z) = \breve{\xi}^{\mathrm{int}}_{z,\pm ,\tilde{n}} (r) \mathcal{H}_{\tilde{n}-1} (z/H),
\label{eq: Kato28} 
\end{equation} 
where $\tilde{n} = n + 1$ or $n-1$ for the coupling with the warp. 

The nonlinear coupling terms are separated into terms proportional to $\exp [i(\omega t -\tilde{m}\varphi)]$ and $\mathcal{H}_{\tilde{n}}(z/H)$. In the case of coupling through $\boldsymbol{\xi}^{W}$, we write the coupling terms as 
\begin{equation}
\frac{1}{2} \rho_{0} \left[ \boldsymbol{C}(\boldsymbol{\xi}, \boldsymbol{\xi}^{W}) + \boldsymbol{C}(\boldsymbol{\xi}^{W}, \boldsymbol{\xi}) \right]_{r} =  \rho_{0} \sum_{\tilde{n}} \breve{A}_{r,+,\tilde{n}} (r) \exp \left[ i(\omega t - \tilde{m}\varphi) \right] \mathcal{H}_{\tilde{n}} (z/H) + \cdots , 
\label{eq: Kato29} 
\end{equation} 
\begin{equation}
\frac{1}{2} \rho_{0} \left[ \boldsymbol{C}(\boldsymbol{\xi}, \boldsymbol{\xi}^{W}) + \boldsymbol{C}(\boldsymbol{\xi}^{W}, \boldsymbol{\xi}) \right]_{\varphi} =  \rho_{0} \sum_{\tilde{n}} \breve{A}_{\varphi,+,\tilde{n}} (r) \exp \left[ i(\omega t - \tilde{m}\varphi) \right] \mathcal{H}_{\tilde{n}} (z/H) + \cdots ,  
\label{eq: Kato30} 
\end{equation} 
\begin{equation}
\frac{1}{2} \rho_{0} \left[ \boldsymbol{C}(\boldsymbol{\xi}, \boldsymbol{\xi}^{W}) + \boldsymbol{C}(\boldsymbol{\xi}^{W}, \boldsymbol{\xi}) \right]_{z} =  \rho_{0} \sum_{\tilde{n}} \breve{A}_{z,+,\tilde{n}} (r) \exp \left[ i(\omega t - \tilde{m}\varphi) \right] \mathcal{H}_{\tilde{n}-1} (z/H) + \cdots ,  
\label{eq: Kato31} 
\end{equation} 
where $\tilde{m} = m+1$ and $+\cdots$ denotes terms orthogonal 
to both $\mathcal{H}_{\tilde{n}}$ and $\mathcal{H}_{\tilde{n}-1}$. 
The subscript $+$ of $\breve{A}$'s denotes that they are related to the $\varphi$-dependence of $\exp [ -i(m+1)\varphi]$. 
Note that these equations are the same as equations (29) -- (31) of \citet{Kat08b}. 
In the case of coupling through $\boldsymbol{\xi}^{W\ast}$, the nonlinear coupling terms, $(1/2) \rho_{0} [ \boldsymbol{C}(\boldsymbol{\xi},\boldsymbol{\xi}^{W\ast})+\boldsymbol{C}(\boldsymbol{\xi}^{W\ast},\boldsymbol{\xi})]$, can be expressed in similar forms, by introducing $\breve{A}_{r,-,\tilde{n}}$,$ \breve{A}_{\varphi ,-,\tilde{n}}$, and $\breve{A}_{z,-,\tilde{n}}$ instead. 
Then, the basic equations describing intermediate oscillations are written as 
\begin{eqnarray}
&\left[  -(\omega - m\Omega)^{2} + \kappa^{2} -4\Omega^{2} -c_{s}^{2} \frac{d^{2}}{dr^{2}} \right] & \breve{\xi}_{r,\pm, \tilde{n}}^{\mathrm{int}} - i2\Omega (\omega - \tilde{m}\Omega) \breve{\xi}_{\varphi ,\pm, \tilde{n}}^{\mathrm{int}} \nonumber \\
&& +  \Omega_{\perp}^{2} H \frac{d\breve{\xi}_{z,\pm, \tilde{n}}^{\mathrm{int}}}{dr} = \breve{A}_{r,\pm,\tilde{n}} ,
\label{eq: Kato32} 
\end{eqnarray} 
\begin{equation}
-(\omega - \tilde{m}\Omega)^{2} \breve{\xi}_{\varphi ,\pm, \tilde{n}}^{\mathrm{int}}+ i2\Omega (\omega - \tilde{m}\Omega) \breve{\xi}_{r,\pm, \tilde{n}}^{\mathrm{int}} = \breve{A}_{\varphi ,\pm,\tilde{n}} ,
\label{eq: Kato33} 
\end{equation} 
\begin{equation}
\left[ -(\omega - \tilde{m}\Omega)^{2} + \tilde{n}\Omega_{\perp}^{2} \right] \breve{\xi}_{z,\pm, \tilde{n}}^{\mathrm{int}} - \tilde{n}\Omega_{\perp}^{2} H \frac{d\breve{\xi}_{r,\pm, \tilde{n}}^{\mathrm{int}}}{dr} = \breve{A}_{z,\pm,\tilde{n}},
\label{eq: Kato34} 
\end{equation} 
where both cases of ${\tilde m}=m+1$ and ${\tilde m}=m-1$ are written together.

We now eliminate $\breve{\xi}_{\varphi,\pm,{\tilde n}}^{\rm int}$ and $\breve{\xi}_{z,\pm,{\tilde n}}^{\rm int}$
from the set of equations (\ref{eq: Kato32}) -- (\ref{eq: Kato34}) to derive an equation with respect to $\breve{\xi}_{r,\pm,{\tilde n}}^{\rm int}$.
By using the approximation that the radial wavelength of oscillations, $\lambda$, is shorter than $r$, 
i.e., $\lambda\ll r$, which has been used in deriving equations (\ref{eq: Kato23}) -- (\ref{eq: Kato25}), we have 
\begin{eqnarray}
 &&\left[ - (\omega -\tilde{m} \Omega )^{2} + \kappa^{2} \right] \breve{\xi}^{\mathrm{int}}_{r,\pm ,\tilde{n}} 
 - \frac{c_{s}^{2}(\omega-{\tilde m}\Omega)^2}{(\omega-{\tilde m}\Omega)^2-{\tilde n}\Omega_\bot^2}
     \frac{d^{2}\breve{\xi}^{\mathrm{int}}_{r,\pm ,\tilde{n}}}{dr^{2}} \nonumber  \\
 &&    = \breve{A}_{r,\pm ,\tilde{n}} - i \frac{2\Omega}{\omega - \tilde{m} \Omega} \breve{A}_{\varphi,\pm ,\tilde{n}}
        +\frac{\Omega_\bot^2H}{(\omega-{\tilde m}\Omega)^2-{\tilde n}\Omega_\bot^2}\frac{d{\breve A}_{z,\pm,{\tilde n}}}{dr}.
\label{eq:Kato35} 
\end{eqnarray} 
The horizontal resonance occurs at the radius where 
$-(\omega-{\tilde m}\Omega)^2+\kappa^2=0$ holds.
After obtaining $\breve{\xi}^{\mathrm{int}}_{r,\pm ,\tilde{n}}$ by solving this equation, we can obtain
$\breve{\xi}^{\mathrm{int}}_{\varphi,\pm ,\tilde{n}}$ and $\breve{\xi}^{\mathrm{int}}_{z,\pm ,\tilde{n}}$ 
from equations (\ref{eq: Kato33}) and (\ref{eq: Kato34}), respectively.

It is noted that in the coupling through ${\tilde n}=n-1$ in warped disks ($n=1$), which is the case 
to be considered below, 
we have $\breve{A}_{z,\pm ,\tilde{n}}=0$
(see the next subsection) and $\breve{\xi}^{\mathrm{int}}_{z,\pm ,\tilde{n}}=0$.

\subsection{Expressions for Coupling Terms}

Detailed expressions for the coupling terms, i.e., the right-hand side of equations (\ref{eq: Kato32}) -- (\ref{eq: Kato34}), are very complicated
(see Appendix for details).
Their expressions, however, can be simplified by using the approximations of $H\ll \lambda \ll r$.
In the case of ${\tilde n}=n-1$, after a lengthy manipulation we have
\begin{eqnarray}
   \breve{A}_{r,\pm ,n-1}= \left[ - \frac{d\Omega_{\bot}^{2}}{dr} \breve{\xi}_{z,n} - \Omega_{\bot}^{2} H  \frac{d}{dr} \left(  \breve{\xi}_{r,n} \frac{d}{dr} \right) + (n-1) \Omega_{\bot}^{2} \breve{\xi}_{z,n} \frac{d}{dr} \right] \left( \begin{array} {cc} \breve{\xi}_{z}^{W} \\ \breve{\xi}_{z}^{W\ast} \end{array} \right),
   \label{eq:Ar+}
\end{eqnarray}
\begin{eqnarray}
   \breve{A}_{\varphi,\pm ,n-1}=  \left[ \pm i n \Omega_{\bot}^{2} \frac{H}{r} \left( \frac{d\breve{\xi}_{r,n}}{dr} + \breve{\xi}_{r,n}  \frac{d}{dr} \right) \mp i \frac{n-1}{r} \Omega_{\bot}^{2} \breve{\xi}_{z,n}  \right] \left( \begin{array} {cc} \breve{\xi}_{z}^{W} \\ \breve{\xi}_{z}^{W\ast} \end{array} \right) ,  
   \label{eq:Aphi+}
\end{eqnarray}
\begin{eqnarray}
   \breve{A}_{z,\pm ,n-1} = \mathrm{small} .                     
   \label{eq:Az+}
\end{eqnarray}
Here, $\breve{A}_{z,\pm,n-1}$ is small in the sense that $H (d\breve{A}_{z,\pm,n-1}) / dr$ is negligible compared with $\breve{A}_{r,\pm,n-1}$ in magnitude. 

Similarly, in the case of ${\tilde n}=n+1$, we have
\begin{eqnarray}
 \breve{A}_{r,\pm ,n+1} = \left[ \Omega_{\bot}^{2} H \frac{d \breve{\xi}_{r,n}}{dr} \frac{d}{dr} \mp \frac{i}{r} \Omega_{\bot}^{2} H \frac{d \breve{\xi}_{\varphi ,n}}{dr}\right] \left( \begin{array} {cc} \breve{\xi}_{z}^{W} \\ \breve{\xi}_{z}^{W\ast} \end{array} \right) , 
   \label{eq:Ar-}
\end{eqnarray}
\begin{equation}
\breve{A}_{\varphi,\pm ,n+1} = \mathrm{smaller \  than\ } \breve{A}_{r, \pm, n+1} \mathrm{\ in \ magnitude}
  \label{eq:Aphi-}
\end{equation}

\begin{eqnarray}
\breve{A}_{z,\pm ,n+1}= \left[ - \frac{d\Omega_{\bot}^{2}}{dr} \breve{\xi}_{r,n} + n \Omega_{\bot}^{2} \breve{\xi}_{r,n} \frac{d}{dr} \mp i n \frac{\Omega_{\bot}^{2}}{r} \breve{\xi}_{\varphi ,n} \right] \left( \begin{array} {cc} \breve{\xi}_{z}^{W} \\ \breve{\xi}_{z}^{W\ast} \end{array} \right).
   \label{eq:Az-}
\end{eqnarray}

It is noted that in the above expressions, ${\breve \xi}_z^{\rm W}$ is adopted for 
$\breve {\mbox{\boldmath $A$}}_{+,{\tilde n}}$, while ${\breve \xi}_z^{\rm W *}$ is adopted 
in the case of $\breve {\mbox{\boldmath $A$}}_{-,{\tilde n}}$, where the asterisk denotes the complex conjugate.

\subsection{Growth Rate}

By the feedback process through the intermediate oscillations, the original oscillation is amplified or damped, 
so the frequency $\omega$ can no longer be real.  
\citet{Kat08b} showed that the growth rate $-\omega_{\rm i}$, where $\omega_{\rm i}$ is the imaginary part of $\omega$, can be written as  

\begin{equation}                 
-\omega_{{\rm i},\pm } = \frac{W_{\pm }}{2E},
\label{eq: Kato12} 
\end{equation}
where $E$ is the wave energy of the original oscillation, $W_{\pm}$ is the rate at which work is done on the original 
oscillation by the nonlinear resonant process, and $\pm$ denotes the cases of the coupling through 
$\breve{\mbox{\boldmath $\xi $}}_{+}^{\rm int}$ and $\breve{\mbox{\boldmath $\xi $}}_{-}^{\rm int}$, respectively. 
Here, $W_{\pm}$ are written as follows, 

\begin{eqnarray}
W_{+}&=\frac{\omega_{0}}{2} \Im \int \frac{1}{2} \rho_{0} \hat{\boldsymbol{\xi}}^{\ast}& \left[ \boldsymbol{C}(\hat{\boldsymbol{\xi}}_{+}^{\rm int},\hat{\boldsymbol{\xi}}^{W\ast}) + \boldsymbol{C}(\hat{\boldsymbol{\xi}}^{W\ast}, \hat{\boldsymbol{\xi}}_{+}^{\rm int}) \right] dV, \nonumber\\
&= \frac{\omega_{0}}{2}\Im\int\rho_{00}(r)&[(2\pi)^{3/2}\tilde{n}!rH] \nonumber \\
&& \times \left[\breve{\xi}_{r,+,\tilde{n}}^{\rm int}\breve{A}_{r,+,\tilde{n}}^{\ast}+ \breve{\xi}_{\varphi ,+,\tilde{n}}^{\rm int}\breve{A}_{\varphi ,+,\tilde{n}}^{\ast} + \frac{1}{\tilde{n}} \breve{\xi}^{\rm int}_{z,+,\tilde{n}} \breve{A}_{z ,+,\tilde{n}}^{\ast} \right]dr,  
\label{eq: Kato1336}
\end{eqnarray} 
 
\begin{eqnarray}
W_{-}&= \frac{\omega_{0}}{2} \Im \int \frac{1}{2} \rho_{0} \hat{\boldsymbol{\xi}}^{\ast} & \left[ \boldsymbol{C}(\hat{\boldsymbol{\xi}}_{-}^{\rm int},\hat{\boldsymbol{\xi}}^{W}) + \boldsymbol{C}(\hat{\boldsymbol{\xi}}^{W}, \hat{\boldsymbol{\xi}}_{-}^{\rm int}) \right]dV \nonumber\\
 & =  \frac{\omega_{0}}{2}\Im\int\rho_{00}(r) & [(2\pi)^{3/2}\tilde{n}!rH]\nonumber \\
&&  \times \left[\breve{\xi}_{r,-,\tilde{n}}^{\rm int}\breve{A}_{r,-,\tilde{n}}^{\ast}+\breve{\xi}_{\varphi ,-,\tilde{n}}^{\rm int}\breve{A}_{\varphi ,-,\tilde{n}}^{\ast} + \frac{1}{\tilde{n}} \breve{\xi}^{\rm int}_{z,-,\tilde{n}} \breve{A}_{z ,-,\tilde{n}}^{\ast}\right]dr,
\label{eq: Kato1437}
\end{eqnarray} 
where $\omega_{0}$ is the frequency of the original oscillation before the mode couplings, and the wave energy, $E$, is given by 
\begin{eqnarray}                 
E &=& \frac{1}{2} \omega_{0} \int \rho_{0} \hat{\xi}^{\ast} \left[ \omega - i(\boldsymbol{u \cdot \nabla}) \right] \hat{\xi}dV \nonumber \\
  && = \frac{(2\pi)^{3/2}}{2} \omega_{0}^{2} (r^{4} H \rho_{00})_{c} E_{n},
\label{eq: Kato1538}  
\end{eqnarray}
where $E_{n}$ is a dimensionless quantity given by 
\begin{equation}                 
E_{n} = \int \frac{ rH \rho_{00}}{(rH \rho_{00})_{c} } \frac{\omega_{0} -m\Omega }{\omega_{0} } \left( n! \frac{\mid \breve{\xi}_{r,n}\mid^{2}}{r_{c}^{3}} + (n-1)! \frac{\mid \breve{\xi}_{z,n}\mid^{2}}{r_{c}^{3}} \right) dr  
\label{eq: Kato39}
\end{equation}
with subscript c denoting the values at the resonant radius, where $-(\omega_{0}-\tilde{m}\Omega)^{2} + \kappa^{2} = 0$. 

\section{Results}
\subsection{Original Oscillations}
As the original oscillations, we consider the axisymmetric g-mode oscillations trapped around the 
radius of the maximum epicyclic frequency, $\kappa_{\rm max}$.
This type of trapped g-mode oscillations has been examined in previous studies (\cite{Oka87}; see also \cite{Kat01},  
\cite{Fer08}).
For this purpose we solve equation (\ref{eq:origosc}) with $m=0$ and $n=1$. 
In what follows, we assume that the sound speed ($c_{\rm s}$, in units of $c$) is constant. 

As mentioned previously, we adopt the radius of ISCO as the inner boundary radius. As the outer boundary radius, we take a radius in the evanescence region of the original oscillations.
The amplitude of the oscillations is
assumed to vanish at both inner and outer boundaries. 

\begin{figure}
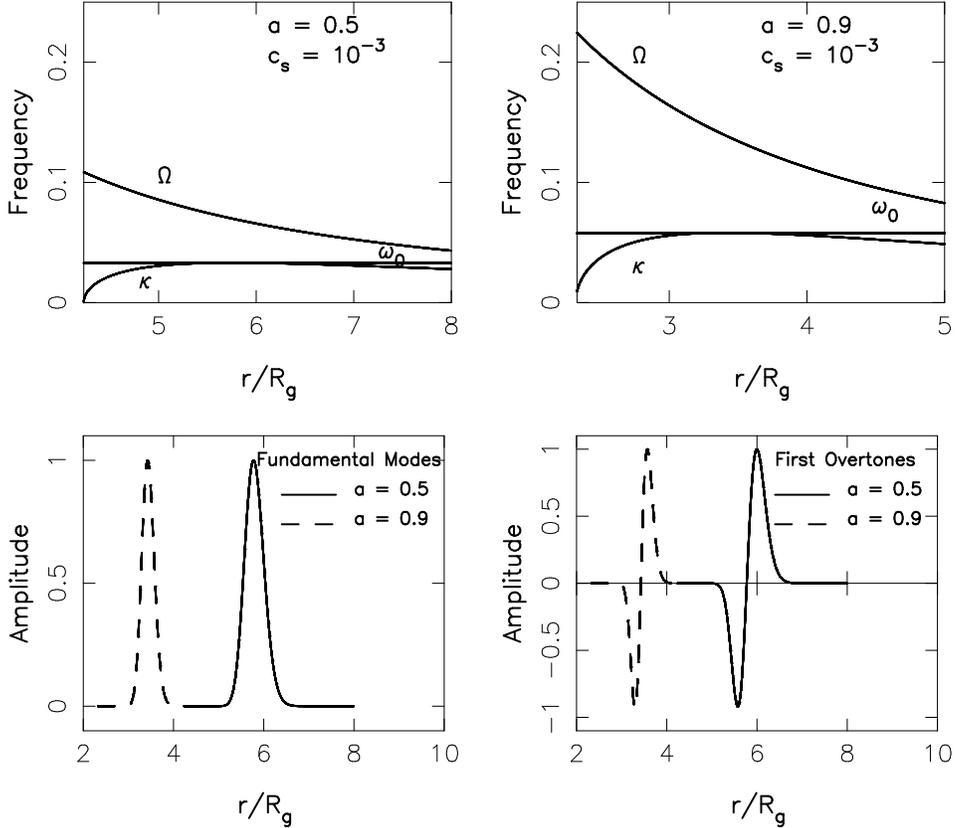

       \centering
         \begin{tabular}{c}
	       \FigureFile(6cm,6cm){propdi5.ps}\\
	       \FigureFile(6cm,6cm){oriosca59.ps}
		\end{tabular}
        \begin{tabular}{c}
		  \FigureFile(6cm,6cm){propdi9.ps}\\
		  \FigureFile(6cm,6cm){orioscfa59.ps}
	   \end{tabular}
	       
		\caption{Eigenmodes and propagation diagrams of the original oscillations for $c_{\rm s}=10^{-3}$. 
                The upper panels are propagation diagrams showing the lowest two eigenmodes for $a=0.5$ (left panel) and $a=0.9$ (right panel). In the lower panels, the eigenfunctions are shown for the fundamental modes (left panel) and the first overtones (right panel) of the original oscillations.}
    \label{fig:oriosc}
\end{figure}

Figure \ref{fig:oriosc} shows the original oscillations for $c_{\rm s}=10^{-3}$. 
The upper panels shows the propagation diagram for the lowest two eigenmodes, the fundamental mode and the
first overtone, although the two horizontal lines showing eigenfrequencies are indistinguishable on the figure. The left panel is for spin parameter $a=0.5$, while the right panel is for $a=0.9$.  Note that the eigenfrequency is higher for a more rapidly rotating system, because the epicyclic frequency, $\kappa$, increases with increasing spin parameter $a$. 
The eigenfrequency is also higher for a lower sound speed. The lower panels show the eigenfunctions of the fundamental modes (left panel) and the first overtones (right panel). The eigenmodes for the two different values of $a$ look similar, but as $a$ increases, the propagation region is shifted closer to the central object.

\subsection{Intermediate Oscillations}

To obtain the intermediate oscillations, we solve equation (\ref{eq:Kato35}) with the nonlinear coupling terms. 
As discussed in the previous section, the azimuthal mode number of the intermediate oscillation is given by 
$\tilde{m}=m\pm 1 = \pm 1$, and the vertical mode number $\tilde{n}=n\pm 1=0$ or $2$. 
By analysing the resonance condition for modes with $Re(\omega )=\omega_{0} \approx \kappa_{\mathrm{max}}$, we find that for $\tilde{m}=-1$ and $\tilde{n}=0$ or $2$, there is no resonance radius 
inside the propagation region of the original oscillation, so that there is no mode excitation in these cases.
Therefore, for the intermediate oscillations, we have two cases of $(\tilde{m}, \tilde{n})=(1,0)$ and $(1,2)$ to study. 

In the case of $({\tilde m}, {\tilde n})=(1,0)$, the intermediate oscillations are p-mode oscillations, 
and their propagation regions are those given by $\omega_{0}\leq \Omega-\kappa$ and $\omega_{0}\geq \Omega+\kappa$.
The intermediate oscillations propagating in the region of $\omega_{0}\leq \Omega-\kappa$ have a resonance at
the outer edge of the region, and the resonant radius is inside or close to the propagation region of the
trapped original oscillations, since $\omega_{0}\sim \kappa_{\rm max}$.
Hence, we consider the intermediate oscillations propagating in the region of $\omega_{0}\leq\Omega-\kappa$.
The boundary conditions to be imposed to the intermediate oscillations are thus vanishing Lagrangian perturbation 
of pressure at the radius of ISCO and zero amplitude 
at the same radius where  
the outer boundary condition is imposed to the original oscillations (see subsequent paragraphs for implications of the latter boundary condition).

In the case of $({\tilde m}, {\tilde n})=(1,2)$, the interesting intermediate oscillations are g-mode oscillations
and their propagation region is $\Omega-\kappa\leq \omega_{0} \leq \Omega+\kappa$.
At the inner edge of the propagation region, we have a resonance of $\omega_{0}=\Omega-\kappa$ and
the resonant radius is inside or close to the trapped region of the original oscillations.
When we impose the outer-boundary condition for the intermediate oscillations, caution is necessary,
since the corotation radius exists inside the propagation region.

At the corotation resonance, g-mode oscillations are damped (\cite{Kat03}, \cite{Li03}).
Hence, this effect should be taken into account.
However, a detailed consideration of the damping effect at the corotation resonance is complicated.
Thus, a practical way is
to introduce a damping of intermediate oscillations at the rate of $\beta\Omega$ with $\beta$ being the damping parameter, as \citet{Fer08} did.

It is noted that unlike the case of $({\tilde m},{\tilde n})=(1,2)$, the effect of corotation resonance 
on the wave behavior will be unimportant in the case of $({\tilde m},{\tilde n})=(1,0)$ for the following reasons.
In the case of p-mode oscillations, the corotation resonance can amplify the oscillations, 
different from the case of g-mode oscillations (\cite{Tsa09}).
Since the corotation radius is inside the evanescent region of the p-mode oscillations, 
the growth rate by corotation resonance depends on the efficiency of penetration of the oscillations through the
evanescent region.
In geometrically thin disks, the efficiency of tunnelling through the evanescent region is low (remember why the Papaloizou-Pringle instability does not have large 
growth rate in geometrically thin disks).
Furthermore, in the case of ${\tilde m}=1$, the evanescent region is wide compared with other cases of ${\tilde m}\geq 2$
(e.g., see Fig.4 of \citet{Tsa09}).
Indeed, \citet{Tsa09} showed that the growth rate in the case of one-armed oscillations ($\tilde{m}=1$) is much 
smaller than that in other cases. Thus, the effect of corotation resonance in the case of ($\tilde{m}, \tilde{n}$) = (1,0) can be neglected in our present problem.

Considering these situations, we focus our attention in this paper only on the case of $({\tilde m},{\tilde n})=(1,0)$,
with no consideration of the effects of corotation resonance, since in this case we can study the resonant 
excitation processes in an idealized form without introducing ambiguities related to the corotation resonance.
 
\begin{figure}
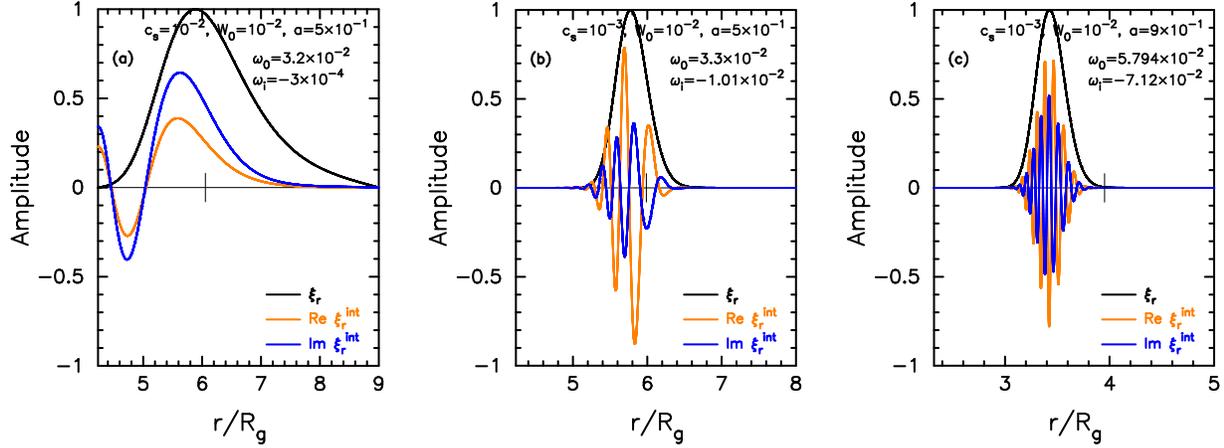

	   \centering
         \begin{tabular}{c}
	       \FigureFile(5cm,6cm){int-eigfunn5camb.ps}
           \end{tabular} 
	      \begin{tabular}{c}
	       \FigureFile(5cm,6cm){int-eigfunn5aamb.ps}
           \end{tabular}     
           \begin{tabular}{c} 
	       \FigureFile(5cm,6cm){int-eigfunn9aamb.ps}
	       \end{tabular}
        	\caption{Eigenfunctions 
        of the fundamental modes of original oscillations and the intermediate oscillations obtained for $W_{0}=0.01$. 
        Panel (a) is for $c_{\rm s}=10^{-2}$ and $a=0.5$, panel (b) is for  
        $c_{\rm s}=10^{-3}$ and $a=0.5$, and panel (c) for $c_{\rm s}=10^{-3}$ and $a=0.9$. 
        The black curve
        is the eigenfunction of the original oscillation, the red and blue curves are
        respectively the real and imaginary parts of the intermediate oscillation. The vertical line denotes the resonance radius of $\omega_{0}=\Omega-\kappa$.}
    \label{fig:intosc}
\end{figure}

Figure \ref{fig:intosc} shows the eigenfunctions of intermediate oscillations with $(\tilde{m}, 
\tilde{n})=(1,0)$ for $c_{\rm s}=10^{-3}$ and $W_{0}=0.01$. 
Three cases of different sets of ($c_{\rm s}$, $a$) are shown.
From the left panel, the adopted set is $(c_{s},a)=(10^{-2}$, $0.5$), ($10^{-3}$, $0.5$), and ($10^{-3}$, $0.9$). 
In each panel, the real and imaginary parts of the intermediate oscillation 
are denoted by the red and blue curves, respectively. The corresponding original oscillation (the fundamental mode) is also shown by the black curve in each panel. 
The vertical line on the $\textit{y}$-axis denotes the resonance radius. Note that as $c_{\rm s}$ decreases or $a$ increases, the intermediate oscillations oscillate with shorter radial wavelength and confined to a narrower region around the resonance radius). Also note that the amplitude of the intermediate oscillations depend little on $c_{\rm s}$ and $a$ (see discussions in the subsequent subsection).

\subsection{Growth Rates}

To calculate the growth rate, $-\omega_{\rm i}$, we need iterative processes.
We first assume a trial value of $-\omega_{\rm i}$, and solve the wave equation (\ref{eq:Kato35}) to obtain
${\breve \xi}_{r,+,\tilde{n}}^{\rm int}$ and ${\breve \xi}_{\varphi ,+,\tilde{n}}^{\rm int}$.
Then, by using equation (\ref{eq: Kato12}), we calculate the growth rate, $-\omega_{\rm i}$.
If the resulting value of $-\omega_{\rm i}$ is not equal to the assumed value, we assume another trial value of $-\omega_{\rm i}$ and repeat this procedure until both values coincide.
Note that in the present case of $(m,n)=(0,1)$ and $({\tilde m}, {\tilde n})=(1,0)$, the work integral, $W_+$, given by 
equation (\ref{eq: Kato1336}) is written explicitly in the form of\footnote{
We sould not confuse  $W_+$ with the warp amplitude $W$.
} 
\begin{eqnarray}
W_{+} = \frac{\omega_{0}}{2} \int & \rho_{00} &(r)[(2\pi)^{3/2}rH] \nonumber \\
&& \times  \breve{\xi}^{\rm int}_{r,{\rm r}} \left\{W r \frac{d \Omega_{\bot}^{2}}{dr}\breve{\xi}_{z} + \Omega_{\bot}^{2} W H \frac{d\breve{\xi}_{r}}{dr}\left(1-\frac{2\Omega}{\omega-\Omega}\right) - \frac{2\Omega}{\omega-\Omega}\Omega_{\bot}^{2} W H \frac{\breve{\xi}_{r}}{r}\right. \nonumber \\
&& \left. + \Omega_{\bot}^{2} H \frac{dW}{dr} \left(r \frac{d\breve{\xi}_{r}}{dr}- \frac{2\Omega}{\omega-\Omega}\breve{\xi}_{r}+2\breve{\xi}_{r}\right) +  \Omega_{\bot}^{2} H r \breve{\xi}_{r} \frac{d^{2}W}{dr^{2}}\right\} dr,
\label{eq:W+exp1}
\end{eqnarray}
where and hereafter $\breve{\xi}^{\rm int}_{r,{\rm r}}$, $\breve{\xi}_{r}$, and $\breve{\xi}_z$ are used to abbreviate $\Re(\breve{\xi}^{\rm int}_{r,+,0})$, $\breve{\xi}_{r,1}$, and $\breve{\xi}_{z,1}$, respectively.

\begin{figure}
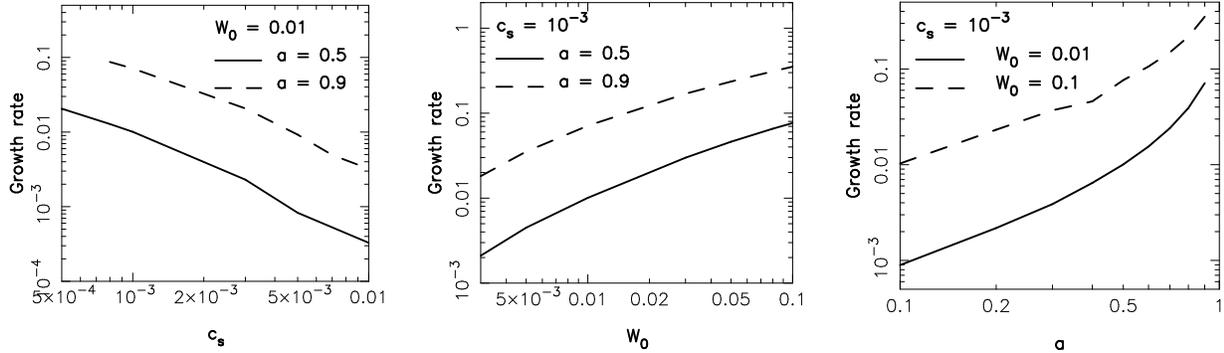

 \centering
         \begin{tabular}{c}
	       \FigureFile(5cm,6cm){mncgwgrcsl.ps}
           \end{tabular} 
	      \begin{tabular}{c} 
	       \FigureFile(5cm,6cm){mncgwgrw0l.ps}
		\end{tabular}
		\begin{tabular}{c}
	       \FigureFile(5cm,6cm){mncgwgral.ps}
           \end{tabular}
    
  	\caption{Variation of the growth rates of the oscillations with the disk structure and spin parameter. 
     The left panel shows the dependence of the growth rate on the sound speed for a fixed warp amplitude at the inner radius, $W_{0}= 0.01$, for $a=0.5$ and $a=0.9$, and the middle panel shows the dependence on $W_{0}$ for a fixed sound speed $c_{\rm s}=10^{-3}$, also for $a=0.5$ and $a=0.9$, while the right panel shows the dependence on the spin parameter, $a$, for a fixed sound speed $c_{\rm s}=10^{-3}$, with $W_{0}= 0.01$ and $W_{0}= 0.1$}
    \label{fig:growth}
\end{figure}

To examine the parameter dependence of the growth rate, however, the following expression for $W_+$ is more perspective than 
equation (\ref{eq:W+exp1}).
Let us multiply the complex conjugates of equations (\ref{eq: Kato32}) and (\ref{eq: Kato33}) by $\breve{\xi}_r^{\rm int}$ and $\breve{\xi}_\varphi^{\rm int}$, respectively, and sum them.
Then, we have $\breve{\xi}_r^{\rm int} A_r^*  +\breve{\xi}_\varphi^{\rm int}A_\varphi^*$ expressed in terms of 
$\breve{\mbox{\boldmath $\xi $}}^{\rm int}$ (and $ \breve{ \mbox{ \boldmath $ \xi $}}^{\rm int *} $ ) alone.
Taking the imaginary part of the expression, we have a simple expression for $W_+$, which is
\begin{eqnarray}
W_{+} = \frac{\omega_{0}}{2} \int & \rho_{00} &(r)[(2\pi)^{3/2}rH] \nonumber \\
&& \times 2\omega_{\rm i}\biggr[(\omega_0-\Omega)(\vert\breve{\xi}_r^{\rm int}\vert^2+\vert\breve{\xi}_\varphi^{\rm int}\vert^2)
     -i\Omega(\breve{\xi}_r^{\rm int}\breve{\xi}_\varphi^{\rm int *}-\breve{\xi}_r^{\rm int *}\breve{\xi}_\varphi^{\rm int})\biggr] dr. 
\label{eq:W+exp1add}
\end{eqnarray}
If this expression for $W_+$ is used, equation (\ref{eq: Kato12}), which can be used to determine the growth rate is  written as
\begin{eqnarray}
   \frac{\omega_{0}}{2} \int & \rho_{00} &(r)[(2\pi)^{3/2}rH] \nonumber \\
&& \times \biggr[(\Omega-\omega_0)(\vert\breve{\xi}_r^{\rm int}\vert^2+\vert\breve{\xi}_\varphi^{\rm int}\vert^2)
     +i\Omega(\breve{\xi}_r^{\rm int}\breve{\xi}_\varphi^{\rm int *}-\breve{\xi}_r^{\rm int *}\breve{\xi}_\varphi^{\rm int})\biggr] dr=E.
\label{eq:W+exp1add2}
\end{eqnarray}
In this equation $\omega_{\rm i}$ does not appear explicitly.
However, $\omega_{\rm i}$ is implicitly involved in $\mbox{\boldmath $\breve{\xi}$}^{\rm int}$, since
$\mbox{\boldmath $\breve{\xi}$}^{\rm int}$ depends on $\omega_{\rm i}$ through  equations (\ref{eq: Kato32}) and (\ref{eq: Kato33}).
Equation (\ref{eq:W+exp1add2}) is further simplified if $\breve{A}_\varphi$ can be neglected compared with
$\breve{A}_r$, which is really allowed if the radial variations of $\breve{\xi}_r$ and $\breve{\xi}_z^{\rm W}$ are rapid
[see equations (\ref{eq:Ar+}) and (\ref{eq:Aphi+})].
In this case, we can adopt $-(\omega_{0}-\Omega)\breve{\xi}_\varphi^{\rm int} 
+ i2\Omega\breve{\xi}_r^{\rm int}\sim 0$, and equation (\ref{eq:W+exp1add2}) is reduced to
\begin{eqnarray}
   \frac{\omega_{0}}{2} \int & \rho_{00} &(r)[(2\pi)^{3/2}rH] \times (\Omega-\omega_0)\vert\breve{\xi}_r^{\rm int}\vert^2 dr=E.
\label{eq:W+exp1add3}
\end{eqnarray}

Calculations of the growth rate have been 
performed by using equation (\ref{eq:W+exp1}) for some parameters 
characterizing the disk structure (warp amplitude at the inner radius $W_{0}$ and acoustic speed $c_{\rm s})$ and spin 
parameter, $a$, of the central source.
Results show, as mentioned before, that the fundamental mode of the trapped
oscillations is excited.
Figure \ref{fig:growth} shows the parameter dependence of the growth rate of the fundamental modes.
The left panel shows the dependence on sound speed $c_{s}$ 
for spin parameter $a=0.5$ and $a=0.9$ with a fixed $W_{0}= 0.01$. This $c_{\rm s}$-dependence of the growth rate is mathematically understandable, 
if we consider equation (\ref{eq:W+exp1add2}) or (\ref{eq:W+exp1add3}).
A smaller value of $c_{\rm s}$ gives rise to shorter radial wavelengths of the original 
oscillations [see equation (\ref{eq:origosc})] 
as well as of the warp amplitude oscillations [see equation (\ref{eq:solwarp})].
This means that the coupling terms between the original oscillation and the warp, $\breve{\mbox{\boldmath $A$}}$,
rapidly vary in the radial direction with large amplitude when $c_{\rm s}$ is small, since the coupling has
terms of the derivative of $\breve{\mbox{\boldmath $\xi $}}$ and $\breve{\mbox{\boldmath $\xi $}}^{\rm W}$
with respect to $r$.
Such large amplitude variations of the coupling terms give rise to large amplitude variations of
the intermediate oscillations [see equation (\ref{eq:Kato35})], if the coefficients of 
$\breve{\xi}_r^{\rm int}$ on the left-hand side of equation (\ref{eq:Kato35}) are fixed.
In other words, if the coefficients are fixed,  condition (\ref{eq:W+exp1add2}) [or (\ref{eq:W+exp1add3})] 
cannot be satisfied in the case of small $c_{\rm s}$.
For condition (\ref{eq:W+exp1add2}) [or (\ref{eq:W+exp1add3})] to be satisfied in the case of a smaller $c_{\rm s}$, 
the coefficients of $\breve{\xi}_r^{\rm int}$ in equation (\ref{eq:Kato35}) must increase 
in the case of a smaller $c_{\rm s}$ so that 
the resulting $\breve{\xi}_r^{\rm int}$ decreases and condition (\ref{eq:W+exp1add2}) [or (\ref{eq:W+exp1add3})] can be
satisfied.
This is realized by increasing the imaginary part of $\omega_{\rm i}$.
Notice, for example, that the imaginary part of the left-hand side of equation (\ref{eq:Kato35}) 
involving $\breve{\xi}_{r,{\rm r}}^{\rm int}$ is $-2\omega_{\rm i}(\omega_0-\Omega)\breve{\xi}_{r,{\rm r}}^{\rm int}$.
In summary, in the case of a smaller $c_{\rm s}$, the value of $\vert\omega_{\rm i}\vert$
must increase so that the amplitude of the intermediate oscillations is roughly independent of $c_{\rm s}$
and the condition (\ref{eq:W+exp1add2}) [or (\ref{eq:W+exp1add3})] is satisfied.
Numerical results shown in panels (a) and (b) of figure \ref{fig:intosc} are consistent with this argument.
That is, comparisons of the results in panels (a) and (b) show that the radial wavelength of the intermediate oscillations becomes
 shorter as $c_{\rm s}$ decreases.
The amplitudes of the intermediate oscillations, however, are roughly on the same order in panels (a) and (b), 
which is realized by an increase of $\vert\omega_{\rm i}\vert$ in panel (b) compared with in panel (a). The result that the growth rate increases with a decrease of $c_{\rm s}$ qualitatively agrees with that of \citet{Fer08}, but quantitatively our result slightly differs from \authorcite{Fer08}`s. For instance, for $c_{s} = 10^{-3}$, $W_{0}=0.01$, and $a=0.5$, our model provides $\mid \omega_{i}\mid = 3.3\times 10^{-4}$, while $\mid \omega_{i}\mid = 4.2\times 10^{-4}$ in \citet{Fer08}. This difference is likely to result from the difference between our analytical approach and their numerical one. 

The middle panel of figure \ref{fig:growth} shows that the growth rate increases with increasing warp amplitude. This is natural, since the oscillations are excited by having the interaction with the warped disk. 
\citet{Fer08} discussed that the growth rate should grow with $(dW/dr)^{2} \propto W_{0}^{2}$ for small $W_{0}$, 
because the excitation consists of two processes: i) the interaction giving rise to the intermediate oscillation and 
ii) feeding back to the original oscillation. 
In fact, their numerical results, which are for $W_{0} \leq  0.02$, agree with this argument. Our model, however, provides  significantly flatter dependance of the growth rate on the warp amplitude than that obtained by \citet{Fer08}. This may be explained as follows: From the real and imaginary parts of equation (\ref{eq:Kato35}), we see that
if $\omega_{\rm i}$ was fixed, both the real and imaginary parts of $\breve{\xi}_{r,+,\tilde{n}}^{\rm int}$, i.e., 
$\breve{\xi}_{r,{\rm r}}^{\rm int}$ and $\breve{\xi}_{r,{\rm i}}^{\rm int}$, would be proportional to 
 the coupling terms, which are proportional to the warp amplitude, $W_{0}$.
Then the work integral, $W_{+}$, (and the growth rate) would be proportional to $W_{0}^{2}$. 
As $\omega_{\rm i}$ increases with increasing $W_{0}$, however, the terms proportional to $\omega_{\rm i}$ 
in the coefficients of
the inhomogeneous wave equations for $\breve{\xi}_{r,{\rm r}}^{\rm int}$ and $\breve{\xi}_{r,{\rm i}}^{\rm int}$ 
[i.e., the terms coming from $[(\omega_{0}-{\tilde m}\Omega)^2+\kappa^2]\breve {\xi}^{\rm int}_r$
in wave equation (\ref{eq:Kato35})] would become non-negligible in determining $\breve{\xi}_{r,{\rm r}}^{\rm int}$ and 
$\breve{\xi}_{r,{\rm i}}^{\rm int}$.
This makes $W_{+}$ (and the growth rate) deviate from $W_{+}\propto W_{0}^{2}$. 
Instead, it leads to a dependence flatter  than $W_{0}^{2}$ as shown in the middle panel of figure \ref{fig:growth}.

From the right panel of figure \ref{fig:growth}, 
we can see the dependence of the growth rate on the spin parameter of the black hole for a fixed sound speed ($c_{\rm s}=10^{-3}$). The growth rate increases with increasing spin parameter. This spin dependence of the growth rate will be due to a similar reason as that in the case of $c_{\rm s}$-dependence.
That is, as the spin parameter increases, the warp rapidly oscillates in the radial direction in the inner region of the disk, 
since the difference between $\Omega$ and $\Omega_\bot$ is large there [see equation (\ref{eq:solwarp})].
Let us assume tentatively that $\omega_{\rm i}$ is unchanged from the value of the case of a small spin parameter.
Then, the intermediate oscillations have a large amplitude with a short wavelength,
since the coupling terms have radial derivative of ${\breve \xi}_z^{\rm W}$ [see equation 
(\ref{eq:Ar+}) and (\ref{eq:Aphi+})].
This means that the condition (\ref{eq:W+exp1add2}) [or (\ref{eq:W+exp1add3})] cannot be satisfied.
For the condition to be satisfied in the case of a large spin parameter, $\vert\omega_{\rm i}\vert$ must increase 
so that the amplitude of the intermediate oscillations decreases
until the magnitude of $\breve{\mbox{\boldmath $\xi $}}^{\rm int}$ becomes comparable with that of the 
case of a smaller spin parameter value.
This will be a reason why the growth rate increases with an increase of the spin.
The result regarding the $a$-dependence of growth rate qualitatively agrees with those obtained by 
\citet{Fer08} for the fundamental g-mode oscillations. Note that for $a=0$ case, we found that there is no excitation of the fundamental g-mode oscillation.

\section{Summary and Discussion}

We have numerically studied the excitation of trapped g-mode oscillations 
in warped 
disks around black holes, based on a scenario proposed by 
\authorcite{Kat04}\ (\yearcite{Kat04, Kat08a, Kat08b}). 
We first obtained trapped g-mode oscillations with eigenfrequencies close to the maximum of the horizontal epicyclic frequency. 
Then, we examined whether these modes are excited via the resonant coupling with the warp 
of the disk. 
We have found that the fundamental modes of the trapped g-modes are excited except in the case of non-rotating black holes. We have performed calculations of the growth rate of the fundamental mode for some parameter values such as the warp amplitude in the inner radius, the isothermal sound speed in the disk and the black hole spin parameter. 
We have found that the growth rate increases as the warp amplitude and spin parameter increases or the sound speed decreases. 
The dependence of the growth rate on these parameters qualitatively agrees with that obtained by \citet{Fer08}
for the fundamental mode, 
but quantitatively there are significant differences between the two.
We suspect that these differences are due to differences of the model considered.

Our results presented in this paper show that the excitation of trapped g-mode oscillations by global disk
deformation is rather sensitive to the disk structure,
because the resonant radius given by $\omega_{0}=\Omega-\kappa$ and a characteristic radius of the oscillation
where ${\breve \xi}_{r,n}$ or $\partial{\breve \xi}_{r,n}/\partial r$ has a peak are close. 
We also suppose that the damping (or amplification) of the intermediate oscillations at 
the corotation
resonance is not an essential ingredient for the excitation of disk oscillations in deformed disks.
In the next paragraphs, we discuss the excitation mechanism from a viewpoint of energy flow among oscillations.

Energy flow between original and intermediate oscillations occurs at two aspects of the resonance. 
The first occurs at the stage where the oscillation resulting from nonlinear coupling between 
the original oscillation and the warp acts as a forcing source on the intermediate oscillation. 
The second occurs at the feedback process from the intermediate oscillation to the original 
oscillation where the intermediate oscillation acts as a source to the original oscillation 
after coupling with the warp. The net energy input rate to the original oscillation through 
these coupling processes is given by equation (\ref{eq: Kato12}). The results derived from 
this equation show that in a simplified situation the signs of wave energy of original and intermediate 
oscillations must be opposite for excitation of the original oscillation 
(e.g., \cite{Kat08a}). 
This suggests that a physical cause of resonant excitation of disk oscillations in 
deformed disks is an interaction of two waves with different signs of the wave energy at a 
resonant radius (\authorcite{Kat08a}\ \yearcite{Kat08a} \yearcite{Kat08b}: \cite{Fer08}). 
In the present case, the original oscillation is axisymmetric and has positive 
energy, i.e., $E>0$. Hence, a necessary condition for wave amplification to occur is 
a negative wave energy of the intermediate oscillation. 
The major process of excitation is thus a positive energy flow from the 
intermediate oscillation with negative energy to the original oscillation with positive energy 
at the resonant region. By this energy flow, both original and intermediate oscillations grow.   

The main role of the warp will be a kind of catalyzer for energy transport. 
In equation (\ref{eq: Kato12}), energy exchange between the oscillation and the disk in the resonant 
region will be involved, but it will be subsidiary in the present excitation mechanism. 
If the intermediate oscillation has an interaction with environment and its negative wave energy
is transferred to the environment, the oscillation has a tendency to be damped. 
However, a negative energy will be transported to the intermediate oscillation from the original 
one to replenish the loss. Then, the original oscillation will be excited more strongly than that 
in the case of absence of such negative energy loss from the intermediate oscillation. 
This is the case considered by \citet{Fer08}.

Here, a brief comment is necessary on the procedure adopted in this paper to evaluate the growth rate.
Our procedure is based on the fact that the growth rate can be calculated by using $W_\pm$ given by
equations (\ref{eq: Kato1336}) and (\ref{eq: Kato1437}).
With these equations,
we need not to solve an inhomogeneous wave equation for the original oscillations.
In other words, the inhomogeneous wave equation to be solved by taking into account the resonant effects 
is only that of intermediate oscillations, as we 
have done in this paper.
This is justified by comparing terms in equations (\ref{eq: Kato1336}) [or equation (\ref{eq: Kato1437})] as follows.

Let us, for example, consider $\Im ({\breve \xi}_r^{\rm int} {\breve A}_r^*)$ as a typical term in the
integrand of equation (\ref{eq: Kato1336}) (the subscripts $+$ and ${\tilde n}$ are omitted for simplicity).
The term consists of ${\breve \xi}_{r,{\rm r}}^{\rm int} {\breve A}_{r, {\rm i}}^* +
{\breve \xi}_{r,{\rm i}}^{\rm int} {\breve A}_{r, {\rm r}}^*$.
If the effect of resonant processes on the original oscillation is neglected, 
${\breve A}_r^*$ is purely imaginary, i.e.,  ${\breve A}_{r, {\rm r}}^*=0$
[see equations (\ref{eq:Ar+}) or (\ref{eq:Ar-})].
Then,  $\Im ({\breve \xi}_r^{\rm int} {\breve A}_r^*)={\breve \xi}_{r,{\rm r}}^{\rm int} {\breve A}_{r, {\rm i}}^*$.
This is nothing but the case we have treated in this paper.
If the effect of resonant processes on the original oscillation is taken into account, however, a real part of
${\breve A}_r^*$ appears, i.e., ${\breve \xi}_{r,{\rm i}}^{\rm int}{\breve A}_{r,{\rm r}}^*\not= 0$.
Then, the problem to be considered here is whether the term ${\breve \xi}_{r,{\rm i}}^{\rm int}{\breve A}_{r,{\rm r}}^*$
is of importance in evaluating $W_+$.
The term is, however, found to be smaller than ${\breve \xi}_{r,{\rm r}}^{\rm int}{\breve A}_{r,{\rm i}}^*$ in
magnitude by the following reasons.

${\breve A}_{r,{\rm r}}^*$ is non-zero, but smaller than ${\breve A}_{r,{\rm i}}^*$ in magnitude,
since it is a small correction term of ${\breve A}_r^*$ introduced by the resonant processes.
Hence, the term ${\breve \xi}_{r,{\rm i}}^{\rm int} {\breve A}_{r, {\rm r}}^*$ is negligible compared with 
${\breve \xi}_{r,{\rm r}}^{\rm int} {\breve A}_{r, {\rm i}}^*$, unless  ${\breve \xi}_{r,{\rm i}}^{\rm int}$
is much larger than ${\breve \xi}_{r,{\rm r}}^{\rm int}$ in magnitude.
The results in this paper show that ${\breve \xi}_{r,{\rm i}}^{\rm int}$ and ${\breve \xi}_{r,{\rm r}}^{\rm int}$
are of the same order in the resonant region (see figures~\ref{fig:intosc}).
Therefore, ${\breve \xi}_{r,{\rm i}}^{\rm int} {\breve A}_{r, {\rm r}}^*$ can be 
safely neglected compared with
${\breve \xi}_{r,{\rm r}}^{\rm int} {\breve A}_{r, {\rm i}}^*$.
In conclusion, when we calculate the growth rate by using $W_{\pm}$, 
the inhomogeneous wave equation that should be considered is only that for the intermediate oscillations, as we have done in this paper.

The results of this paper as well as those of \citet{Fer08} confirm that the
resonant excitation process of disk oscillations in deformed disks work to excite trapped oscillations.
The excitation process is thus one of the most
promising causes of the observed high frequency QPOs.
The trapped g-mode oscillations themselves, however, cannot account for the observed
characteristic that high frequency QPOs often appear in pairs with frequency ratio
close to 3:2.
Furthermore, in the case of kHz QPOs in neutron-star low-mass X-ray sources,
the frequencies of QPOs vary with time.
To describe such observational characteristics by the present disk oscillation models,
it might be necessary to relax the restriction of trapping, i.e.,    
consideration of non-trapped oscillations might be necessary, as \authorcite{Kat04} (\yearcite{Kat04}, \yearcite{Kat08a}) did. 

\bigskip

We thank the anonymous referee for invaluable comments, which greatly helped us improve the paper. FO thanks Masayuki Fujimoto for helpful discussions. She also acknowledges the scholarship from Ministry of Education, Culture, Sports, Science and Technology. ATO is grateful for the finantial support via research grants from Hokkai-Gakuen Educational Foundation and Japan Society for the Promotion of Science (20540236).

\appendix
\section{Calculation of Nonlinear Coupling Terms}
The coupling terms used for the calculation is written as 

\begin{equation}
\rho_{0} \boldsymbol{C} (\boldsymbol{\xi},\boldsymbol{\xi} ) = \rho_{0} \boldsymbol{A}^{\psi } 
   + \rho_{0} \boldsymbol{A}^{p},
\label{eq:C1a}
\end{equation} 
where 
\begin{equation}
\rho_{0} \boldsymbol{A}^{\psi } = - \frac{1}{2} \rho_{0} \xi_{i} \xi_{j} \frac{\partial^{2}}{\partial r_{i} \partial r_{j}} \left( \nabla\psi_{0} \right), 
\label{eq:C1b}
\end{equation} 

\begin{equation}
\rho_{0} \boldsymbol{A}^{p} = - \frac{\partial}{\partial r_{i}} \left( p_{0} \frac{\partial\xi_{i}}{\partial r_{j}} \nabla \xi_{j} \right).
\label{eq:C1c}
\end{equation} 
Their detailed expressions for $r$-, $\varphi$-, and $z$-components are given in Appendix 1 of Kato (2004).

For the original oscillation we have 

\begin{equation}
{\xi}_{r} = \exp [i(\omega t - m\varphi)] \mathcal{H}_{n} \breve{\xi}_{r,n},
\label{eq:C2a}
\end{equation} 

\begin{equation}
{\xi}_{\varphi} = \exp [i(\omega t - m\varphi)] \mathcal{H}_{n} \breve{\xi}_{\varphi,n},
\label{eq:C2b}
\end{equation} 

\begin{equation}
{\xi}_{z} = \exp [i(\omega t - m\varphi)] \mathcal{H}_{n-1} \breve{\xi}_{z,n},
\label{eq:C2c}
\end{equation} 
and for warps we have 

\begin{equation}
{\xi}_{r}^{W} = \exp ( - i \varphi ) \mathcal{H}_{1} \breve{\xi}_{r}^{W}, 
\label{eq:C2d}
\end{equation} 

\begin{equation}
{\xi}_{\varphi}^{W}  = \exp ( - i \varphi ) \mathcal{H}_{1} \breve{\xi}_{\varphi}^{W}, 
\label{eq:C2e}
\end{equation} 

\begin{equation}
{\xi}_{z}^{W}  = \exp ( - i \varphi ) \mathcal{H}_{0} \breve{\xi}_{z}^{W}, 
\label{eq:C2f}
\end{equation} 
where $\breve{\xi}_{r}^{W}= -i W(r) H $, $\breve{\xi}_{\varphi}^{W}= -W(r) H$, and $\breve{\xi}_z^{W}=iW(r) r$.

\subsection{Expressions for $\breve{\boldsymbol{A}}^{\psi}$ in Cylindrical Coordinates}

Their derivations from equation (\ref{eq:C1b}) are relatively easy.
Using $O({\breve \xi}_{z,n}) = n(H/\lambda)O({\breve \xi}_{r,n})$ and 
the approximations of $H\ll \lambda \ll r$, we have, after some manipulation,
\begin{eqnarray}
\breve{A}^{\psi }_{r,\pm ,n-1} &=& - n \frac{d^{2}}{dr^{2}}(\Omega^{2} r) \breve{\xi}_{r,n}\left( \begin{array} {cc} \breve{\xi}_{r}^{W} \\ \breve{\xi}_{r}^{W\ast} \end{array} \right) - n \frac{d\Omega_\bot^{2}}{dr} \breve{\xi}_{\varphi,n} \left( \begin{array} {cc} \breve{\xi}_{\varphi}^{W} \\ \breve{\xi}_{\varphi}^{W\ast} \end{array} \right) \nonumber \\
&&- \frac{d\Omega^{2}}{dr}  \breve{\xi}_{z ,n} \left( \begin{array} {cc} \breve{\xi}_{z}^{W} \\ \breve{\xi}_{z}^{W\ast} \end{array} \right),
\label{eq:new11}
\end{eqnarray} 

\begin{eqnarray}
\breve{A}^{\psi }_{r,\pm ,n+1} &=& - \frac{d^{2}}{dr^{2}}(\Omega^{2} r) \breve{\xi}_{r,n} \left( \begin{array} {cc} \breve{\xi}_{r}^{W} \\ \breve{\xi}_{r}^{W\ast} \end{array} \right) - \frac{d\Omega^{2}}{dr} \breve{\xi}_{\varphi ,n} \left( \begin{array} {cc} \breve{\xi}_{\varphi}^{W} \\ \breve{\xi}_{\varphi}^{W\ast} \end{array} \right) \nonumber \\
&& - H \frac{d^{2}\Omega_\bot^{2}}{dr^{2}} \breve{\xi}_{r,n}  \left( \begin{array} {cc} \breve{\xi}_{z}^{W} \\ \breve{\xi}_{z}^{W\ast} \end{array} \right),  
\label{eq:new12}
\end{eqnarray} 

\begin{eqnarray}
\breve{A}^{\psi }_{\varphi,\pm ,n-1} &=& - n H \frac{d\Omega^{2}}{dr} \breve{\xi}_{\varphi ,n}\left( \begin{array} {cc} \breve{\xi}_{r}^{W} \\ \breve{\xi}_{r}^{W\ast} \end{array} \right) - n \frac{d\Omega^{2}}{dr}  \breve{\xi}_{r,n}\left( \begin{array} {cc} \breve{\xi}_{\varphi}^{W} \\ \breve{\xi}_{\varphi}^{W\ast} \end{array} \right) \nonumber \\
&& - n \frac{H}{r}\frac{d\Omega_\bot^{2}}{dr}\breve{\xi}_{\varphi ,n} \left( \begin{array} {cc} \breve{\xi}_{z}^{W} \\ \breve{\xi}_{z}^{W\ast} \end{array} \right),  
\label{eq:new13}
\end{eqnarray} 

\begin{eqnarray}
\breve{A}^{\psi }_{\varphi,\pm ,n+1} &=& - H \frac{d\Omega^{2}}{dr} \breve{\xi}_{\varphi ,n}\left( \begin{array} {cc} \breve{\xi}_{r}^{W} \\ \breve{\xi}_{r}^{W\ast} \end{array} \right) - \frac{d\Omega^{2}}{dr}  \breve{\xi}_{r,n}\left( \begin{array} {cc} \breve{\xi}_{\varphi}^{W} \\ \breve{\xi}_{\varphi}^{W\ast} \end{array} \right) \nonumber \\
&& - \frac{H}{r}\frac{d\Omega_\bot^{2}}{dr}\breve{\xi}_{\varphi ,n} \left( \begin{array} {cc} \breve{\xi}_{z}^{W} \\ \breve{\xi}_{z}^{W\ast} \end{array} \right),  
\label{eq:new14}
\end{eqnarray} 

\begin{eqnarray}
\breve{A}^{\psi }_{z,\pm ,n-1} = -(n-1) \frac{d\Omega_\bot^{2}}{dr} \breve{\xi}_{z,n} \left( \begin{array} {cc} \breve{\xi}_{r}^{W} \\ \breve{\xi}_{r}^{W\ast} \end{array} \right) - 3 (n-1)\frac{H}{r}\frac{d\Omega_\bot^{2}}{dr} \breve{\xi}_{z,n}\left( \begin{array} {cc} \breve{\xi}_{z}^{W} \\ \breve{\xi}_{z}^{W\ast} \end{array} \right),
\label{eq:C55}
\end{eqnarray} 

\begin{eqnarray}
\breve{A}^{\psi }_{z,\pm ,n+1} = -\frac{d\Omega_\bot^{2}}{dr} \breve{\xi}_{r,n} \left( \begin{array} {cc} \breve{\xi}_{z}^{W} \\ \breve{\xi}_{z}^{W\ast} \end{array} \right).  
\label{eq:C55}
\end{eqnarray}

\subsection{Expressions for $\breve{\boldsymbol{A}}^{p}$ in Cylindrical Coordinates}

Derivation of detailed expressions for $\breve{A}_r^p$, $\breve{A}_\varphi^p$, and $\breve{A}_z^p$ are somewhat troublesome.
Below we consider them separately.

\subsubsection{Expression for $\breve{A}_r^p$}

From equation (\ref{eq:C1c}) we have

\begin{equation}
\rho_{0} A_{r}^{p} = - \frac{\partial T_{rr}}{\partial r} -\frac{\partial T_{\varphi r}}{r\partial\varphi}-\frac{\partial T_{zr}}{\partial z}- \frac{1}{r} T_{rr} + \frac{1}{r} T_{\varphi\varphi},
\label{eq:C61}
\end{equation} 
where in the case of the coupling through $\mbox{\boldmath $\xi $}^{\rm int}_{-}$, 
$T_{rr}$, $T_{\varphi r}$, $T_{zr}$, and $T_{\varphi \varphi}$ in equation (\ref{eq:C61}) are respectively given by

\begin{eqnarray}
T_{rr} &=& p_{0} \left[ \frac{\partial\xi_{r}}{\partial r} \frac{\partial\xi_{r}^{W\ast}}{\partial r} + \frac{\partial\xi_{r}^{W\ast}}{\partial r} \frac{\partial\xi_{r}}{\partial r} + \left( \frac{\partial\xi_{r}}{r\partial\varphi} - \frac{\xi_{\varphi}}{r} \right) \frac{\partial\xi_{\varphi}^{W\ast}}{\partial r} \right. \nonumber \\
&& \left. + \left( \frac{\partial\xi_{r}^{W\ast}}{r\partial\varphi} - \frac{\xi_{\varphi}^{W\ast}}{r} \right) \frac{\partial\xi_{\varphi}}{\partial r} + \frac{\partial\xi_{r}}{\partial z} \frac{\partial\xi_{z}^{W\ast}}{\partial r} + \frac{\partial\xi_{r}^{W\ast}}{\partial z} \frac{\partial\xi_{z}}{\partial r} \right],
\label{eq:new21}
\end{eqnarray} 

\begin{eqnarray}
T_{\varphi r} &=& p_{0} \left[ \frac{\partial\xi_{\varphi }}{\partial r} \frac{\partial\xi_{r }^{W\ast}}{\partial r} + \frac{\partial\xi_{\varphi }^{W\ast}}{\partial r} \frac{\partial\xi_{r}}{\partial r} + \left( \frac{\partial\xi_{\varphi }}{r\partial\varphi} + \frac{\xi_{r}}{r} \right) \frac{\partial\xi_{\varphi}^{W\ast}}{\partial r} \right. \nonumber \\
&& \left. + \left( \frac{\partial\xi_{\varphi }^{W\ast}}{r\partial\varphi} + \frac{\xi_{r}^{W\ast}}{r} \right) \frac{\partial\xi_{\varphi}}{\partial r} + \frac{\partial\xi_{\varphi }}{\partial z} \frac{\partial\xi_{z}^{W\ast}}{\partial r} + \frac{\partial\xi_{\varphi }^{W\ast}}{\partial z} \frac{\partial\xi_{z}}{\partial r} \right],
\label{eq:new31}
\end{eqnarray} 

\begin{eqnarray}
T_{zr} &=& p_{0} \left[ \frac{\partial\xi_{z}}{\partial r} \frac{\partial\xi_{r}^{W\ast}}{\partial r} + \frac{\partial\xi_{z}^{W\ast}}{\partial r} \frac{\partial\xi_{r}}{\partial r} + \frac{\partial\xi_{z}}{r\partial\varphi}\frac{\partial \xi_{\varphi}^{W\ast}}{\partial r}+\frac{\partial\xi_{z}^{W\ast}}{r\partial\varphi}\frac{\partial \xi_{\varphi}}{\partial r}\right. \nonumber \\
&& \left. + \frac{\partial\xi_{z}}{\partial z} \frac{\partial\xi_{z}^{W\ast}}{\partial r} + \frac{\partial\xi_{z}^{W\ast}}{\partial z} \frac{\partial\xi_{z}}{\partial r} \right],
\label{eq:new32}
\end{eqnarray} 

\begin{eqnarray}
T_{\varphi\varphi} &=& p_{0} \left[ \frac{\partial\xi_{\varphi}}{\partial r} \left( \frac{\partial\xi_{r}^{W\ast}}{r \partial\varphi} - \frac{\xi_{\varphi}^{W\ast}}{r} \right) + \frac{\partial\xi_{\varphi}^{W\ast}}{\partial r} \left( \frac{\partial\xi_{r}}{r \partial\varphi} - \frac{\xi_{\varphi}}{r} \right) \right. \nonumber \\
&& \left. + 2 \left( \frac{\partial\xi_{\varphi}}{r \partial\varphi}+\frac{\xi_{r}}{r} \right)\left( \frac{\partial\xi_{\varphi}^{W\ast}}{r \partial\varphi}+\frac{\xi_{r}^{W\ast}}{r} \right)\right. \nonumber \\
&& \left. + \frac{\partial\xi_{\varphi}}{\partial z} \frac{\partial\xi_{z}^{W\ast}}{r \partial\varphi} + \frac{\partial\xi_{\varphi}^{W\ast}}{\partial z} \frac{\partial\xi_{z}}{r \partial\varphi} \right].
\label{eq:new33}
\end{eqnarray} 
For coupling through $\mbox{\boldmath $\xi $}^{\rm int}_{+}$, 
every $\mbox{\boldmath $\xi $}^{W\ast}$ in equations (\ref{eq:new21})--(\ref{eq:new33}) should be replaced with $\mbox{\boldmath $\xi $}^{W}$.

Then, taking only the leading order terms under $O(\breve{\xi}_{z,n})=n(H/\lambda)O(\breve{\xi}_{r,n})$ and the approximation of $H\ll \lambda \ll r$, we have 

\begin{eqnarray}
\breve{A}^{p}_{r,\pm,n-1} = \left[ - n \Omega_{\bot}^{2} H  \left( \frac{d\breve{\xi}_{r,n}}{dr} + \breve{\xi}_{r,n} \frac{d}{dr} \right) \frac{d}{dr}+ (n-1) \Omega_{\bot}^{2} \breve{\xi}_{z,n} \frac{d}{dr} \right]\left( \begin{array} {cc} \breve{\xi}_{z}^{W} \\ \breve{\xi}_{z}^{W\ast} \end{array} \right), 
\label{eq:newp91}
\end{eqnarray}

\begin{eqnarray}
\breve{A}^{p}_{r,\pm,n+1} = \left[ \Omega_{\bot}^{2} H  \frac{d\breve{\xi}_{r,n}}{dr} \frac{d}{dr} \mp \frac{i}{r} \Omega_{\bot}^{2} H \frac{d\breve{\xi}_{\varphi,n}}{dr} \right] \left( \begin{array} {cc} \breve{\xi}_{z}^{W} \\ \breve{\xi}_{z}^{W\ast} \end{array} \right). 
\label{eq:newp92}
\end{eqnarray}

\subsubsection{Expression for $\breve{A}_\varphi^p$}

From equation (\ref{eq:C1c}), we have

\begin{eqnarray}
\rho_{0} A_{\varphi}^{p} = - \frac{\partial T_{r\varphi}}{\partial r} -\frac{\partial T_{\varphi\varphi }}{r\partial\varphi}-\frac{\partial T_{z\varphi }}{\partial z}- \frac{1}{r}( T_{r\varphi }+T_{\varphi r}),
\label{eq:newp8}
\end{eqnarray} 
where in the case of the coupling through $\mbox{\boldmath $\xi $}_-^{\rm int}$, 
$T_{r\varphi}$ and $T_{z\varphi}$ are given, respectively, by

\begin{eqnarray}
T_{r\varphi} &=& p_{0} \left[ \left( \frac{\partial\xi_{r}}{\partial r}+ \frac{\partial\xi_{\varphi }}{r \partial \varphi}+\frac{\xi_{r}}{r} \right)\left( \frac{\partial\xi_{r}^{W\ast}}{r \partial\varphi} - \frac{\xi_{\varphi}^{W\ast}}{r} \right) \right. \nonumber \\
&& \left. + \left( \frac{\partial\xi_{r}^{W\ast}}{\partial r}+\frac{\partial\xi_{\varphi }^{W\ast}}{r\partial\varphi}+ \frac{\xi_{r}^{W\ast}}{r} \right)\left( \frac{\partial\xi_{r}}{r \partial\varphi}-\frac{\xi_{\varphi }}{r} \right)\right. \nonumber \\
&& \left. + \frac{\partial\xi_{r}}{\partial z} \frac{\partial\xi_{z}^{W\ast}}{r \partial\varphi} + \frac{\partial\xi_{r}^{W\ast}}{\partial z} \frac{\partial\xi_{z}}{r \partial\varphi} \right],
\label{eq:newp6a}
\end{eqnarray} 

\begin{eqnarray}
T_{z\varphi} &=& p_{0} \left[ \frac{\partial\xi_{z}}{\partial r} \left( \frac{\partial\xi_{r}^{W\ast}}{r \partial\varphi} - \frac{\xi_{\varphi}^{W\ast}}{r} \right) + \frac{\partial\xi_{z}^{W\ast}}{\partial r} \left( \frac{\partial\xi_{r}}{r \partial\varphi} - \frac{\xi_{\varphi}}{r} \right) \right. \nonumber \\
&& \left. + \frac{\partial \xi_{z}}{r\partial\varphi} \left( \frac{\partial\xi_{\varphi}^{W\ast}}{r \partial\varphi}+\frac{ \xi_{r}^{W\ast}}{r} \right)+ \frac{\partial \xi_{z}^{W\ast}}{r\partial\varphi}\left( \frac{\partial\xi_{\varphi}}{r \partial\varphi}+\frac{\xi_{r}}{r} \right)\right. \nonumber \\
&& \left. + \frac{\partial\xi_{z}}{\partial z} \frac{\partial\xi_{z}^{W\ast}}{r \partial\varphi} + \frac{\partial\xi_{z}^{W\ast}}{\partial z} \frac{\partial\xi_{z}}{r \partial\varphi} \right].
\label{eq:newp6b}
\end{eqnarray} 
As in the expressions for $\breve{A}_r^p$, every $\mbox{\boldmath $\xi $}^{W\ast}$ 
in equations (\ref{eq:newp6a}) and (\ref{eq:newp6b}) should be replaced with 
$\mbox{\boldmath $\xi $}^{W}$ in the case of coupling through 
$\mbox{\boldmath $\xi $}^{\rm int}_{+}$.

Substituting equations (\ref{eq:newp6a}) and (\ref{eq:newp6b}) into 
equation (\ref{eq:newp8}), and taking only the leading order terms
as we did before, we have
 
\begin{eqnarray}
\breve{A}^{p}_{\varphi ,\pm ,n-1} = \left[ \pm i  n \frac{n \Omega_\bot^2 H}{r} \left( \frac{d\breve{\xi}_{r,n}}{dr} + \breve{\xi}_{r,n}  \frac{d}{dr} \right) \mp i \frac{(n-1)}{r} \Omega_\bot^2 \breve{\xi}_{z,n} \right] \left( \begin{array} {cc} \breve{\xi}_{z}^{W} \\ \breve{\xi}_{z}^{W\ast} \end{array} \right),
\label{eq:newp10}
\end{eqnarray} 

\begin{eqnarray}
\breve{A}^{p}_{\varphi ,\pm ,n+1}= \left[ - \frac{\Omega_\bot^2 H }{r} \left( i m \breve{\xi}_{r,n} + \breve{\xi}_{\varphi,n} \right) \frac{d}{dr} \pm  \frac{\Omega_\bot^2 H }{r^{2}} \left( - i \breve{\xi}_{r,n} - m \breve{\xi}_{\varphi,n} \right) \right] \left( \begin{array} {cc} \breve{\xi}_{z}^{W} \\ \breve{\xi}_{z}^{W\ast} \end{array} \right).
\label{eq:newp11}
\end{eqnarray}

\subsubsection{Expression for $\breve{A}_z^p$}

The $z$-component of equation (\ref{eq:C1c}) is written as

\begin{equation}
\rho_{0} A_{z}^{p} = - \frac{\partial T_{rz}}{\partial r} -\frac{\partial T_{\varphi z}}{r\partial\varphi}-\frac{\partial T_{zz }}{\partial z}- \frac{1}{r}T_{rz},
\label{eq:Cz22}
\end{equation} 
where in the case of the coupling through $\mbox{\boldmath $\xi $}_-^{\rm int}$,
$T_{rz}$, $T_{\varphi z}$, and $T_{zz}$ are given respectively by
\begin{eqnarray}
T_{rz} &=& p_{0} \left[ \frac{\partial\xi_{r}}{\partial r} \frac{\partial\xi_{r}^{W\ast}}{\partial z} + \frac{\partial\xi_{r}^{W\ast}}{\partial r} \frac{\partial\xi_{r}}{\partial z} + \left( \frac{\partial \xi_{r}}{r \partial \varphi}- \frac{\xi_{\varphi}}{r} \right) \frac{\partial\xi_{\varphi}^{W\ast}}{\partial z} \right. \nonumber \\
&& \left. + \left( \frac{\partial \xi_{r}^{W\ast}}{r \partial \varphi}- \frac{\xi_{\varphi}^{W\ast}}{r} \right) \frac{\partial\xi_{\varphi}}{\partial z} + \frac{\partial\xi_{r}}{\partial z} \frac{\partial\xi_{z}^{W\ast}}{\partial z} + \frac{\partial\xi_{r}^{W\ast}}{\partial z} \frac{\partial\xi_{z}}{\partial z} \right],
\label{eq:Cz23}
\end{eqnarray} 

\begin{eqnarray}
T_{\varphi z} &=& p_{0} \left[ \frac{\partial\xi_{\varphi}}{\partial r} \frac{\partial\xi_{r}^{W\ast}}{\partial z} + \frac{\partial\xi_{\varphi}^{W\ast}}{\partial r} \frac{\partial\xi_{r}}{\partial z} + \left( \frac{\partial \xi_{\varphi}}{r \partial \varphi}+ \frac{\xi_{r}}{r} \right) \frac{\partial\xi_{\varphi}^{W\ast}}{\partial z} \right. \nonumber \\
&& \left. + \left( \frac{\partial \xi_{\varphi}^{W\ast}}{r \partial \varphi}+ \frac{\xi_{r}^{W\ast}}{r} \right) \frac{\partial\xi_{\varphi}}{ \partial z} + \frac{\partial\xi_{\varphi}}{\partial z} \frac{\partial\xi_{z}^{W\ast}}{\partial z} + \frac{\partial\xi_{\varphi}^{W\ast}}{\partial z}  \frac{\partial\xi_{z}}{\partial z} \right],
\label{eq:Cz31}
\end{eqnarray} 

\begin{eqnarray}
T_{z z} &=& p_{0} \left[ \frac{\partial\xi_{z}}{\partial r} \frac{\partial\xi_{r}^{W\ast}}{\partial z} + \frac{\partial\xi_{z}^{W\ast}}{\partial r} \frac{\partial\xi_{r}}{\partial z} + \frac{\partial \xi_{z}}{r \partial \varphi} \frac{\partial \xi_{\varphi}^{W\ast}}{\partial z} + \frac{\partial\xi_{z}^{W\ast}}{r \partial \varphi} \frac{\partial \xi_{\varphi}}{\partial z}+ 2 \frac{\partial \xi_{z}}{\partial z } \frac{\partial\xi_{z}^{W\ast}}{\partial z} \right].
\label{eq:Cz32}
\end{eqnarray} 
For coupling through $\mbox{\boldmath $\xi $}^{\rm int}_{+}$, 
every $\mbox{\boldmath $\xi $}^{W\ast}$ 
in equations (\ref{eq:Cz23})--(\ref{eq:Cz32}) should be replaced with 
$\mbox{\boldmath $\xi $}^{W}$.
Then, with the same procedures taking only the leading order terms
as we did for $\breve{A}_r^p$ and $\breve{A}_\varphi^p$, 
the following expressions are obtained:
\begin{eqnarray}
\breve{A}^{p}_{z ,\pm ,n-1} = \left[ - \Omega_{\bot}^{2} H (n-1) \frac{d}{dr}  \left( n  \breve{\xi}_{r,n} \frac{d}{dr} + \frac{1}{\mathrm{H}} \breve{\xi}_{z,n} \right) \right] \left( \begin{array} {cc} \breve{\xi}_{r}^{W} \\ \breve{\xi}_{r}^{W\ast} \end{array} \right) ,
\label{eq:newp15a}
\end{eqnarray}

\begin{eqnarray}
\breve{A}^{p}_{z ,\pm ,n+1}& = & - \Omega_{\bot}^{2} \mathrm{H} \left( \frac{d^{2}\breve{\xi}_{r,n}}{dr^{2}} +  \frac{d\breve{\xi}_{r,n}}{dr} \frac{d}{dr} \right) \left( \begin{array} {cc} \breve{\xi}_{r}^{W} \\ \breve{\xi}_{r}^{W\ast} \end{array} \right) \nonumber \\
&& + \frac{n \Omega_{\bot}^{2}}{r} \left( \breve{\xi}_{r,n} r \frac{d}{dr} \mp i \breve{\xi}_{\varphi,n} \right) \left( \begin{array} {cc} \breve{\xi}_{z}^{W} \\ \breve{\xi}_{z}^{W\ast} \end{array} \right).
\label{eq:newp15b}
\end{eqnarray}

Using equations (\ref{eq:new11})--(\ref{eq:C55}) and equations (\ref{eq:newp91}), 
(\ref{eq:newp92}), (\ref{eq:newp10}), (\ref{eq:newp11}), (\ref{eq:newp15a}), and (\ref{eq:newp15b}) 
with $O(\breve{\xi}_{z,n})=n(H/\lambda)O\breve{\xi}_{r,n})$ and $H\ll \lambda \ll r$, we finally have equations (\ref{eq:Ar+}) -- (\ref{eq:Az-})
given in the text.

\end{document}